# The Cactus Worm:
# Experiments with Dynamic Resource Discovery and Allocation in a Grid Environment

Gabrielle Allen[♦]   David Angulo[¶]   Ian Foster[•¶]   Gerd Lanfermann[♦]   Chuang Liu[¶]
Thomas Radke[♦]   Ed Seidel[♦]   John Shalf[#]

## Abstract

The ability to harness heterogeneous, dynamically available "Grid" resources is attractive to typically resource-starved computational scientists and engineers, as in principle it can increase, by significant factors, the number of cycles that can be delivered to applications. However, new adaptive application structures and dynamic runtime system mechanisms are required if we are to operate effectively in Grid environments. In order to explore some of these issues in a practical setting, we are developing an experimental framework, called Cactus, that incorporates both adaptive application structures for dealing with changing resource characteristics and adaptive resource selection mechanisms that allow applications to change their resource allocations (e.g., via migration) when performance falls outside specified limits. We describe here the adaptive resource selection mechanisms and describe how they are used to achieve automatic application migration to "better" resources following performance degradation. Our results provide insights into the architectural structures required to support adaptive resource selection. In addition, we suggest that this "Cactus Worm" is an interesting challenge problem for Grid computing.

## 1   Introduction

Dramatic improvements in local and wide area network performance (e.g., Gb/s Ethernet, 10 Gb/s Ethernet, InfiniBand, optical networking) make it increasingly feasible to use collections of networked commodity processors as computational resources. This phenomenon accounts for the popularity of commodity clusters today. We believe that it will also lead, in the near future, to the widespread use of dynamically assembled collections of computers, located on local and even wide area networks, as powerful computational resources. Thus, for example, a computational task might be mapped initially to available processors within a workgroup, but then, as either the characteristics of the computation and/or resource availability change, extend or migrate to other resources available within a physical or "virtual" organization — and/or to resources provided by a commercial computational services provider.

The ability to assemble computational resources in this way should, in principle, be of considerable significance for those many communities (e.g., scientific simulation, engineering design) for whom inadequate access to computation is an obstacle to progress. However, efficient execution in such "Grid" environments can be challenging due to the heterogeneous, dynamic nature of the resources involved. Hence, we find that while successes have been achieved for trivially parallel parameter study problems, it is certainly not routinely possible to run arbitrary parallel programs in a Grid environment.

---

[♦] Max-Planck-Institut für Gravitationsphysik, Albert-Einstein-Institut (AEI), 14476 Golm, Germany
[¶] Department of Computer Science, The University of Chicago, Chicago, IL 60657, USA.
[•] Mathematics and Computer Science Division, Argonne National Laboratory, Argonne, IL 60439, USA.
[#] Lawrence Berkeley National Laboratory, Berkeley, CA 94720, USA.

We believe that a promising approach to this problem is to construct *Grid-enabled computational frameworks* that incorporate the adaptive techniques required for operation in dynamic Grid environments and *Grid runtimes* that provide key services required by such frameworks, such as security, resource discovery, and resource co-allocation. Such computational frameworks and runtimes allow users to code applications at a high level of abstraction (e.g., as operations on multi-dimensional arrays), delegating to the framework and runtime difficult issues relating to distribution across Grid resources, choice of algorithm, and so forth. Such frameworks and runtimes have of course been applied extensively within parallel computing; however, the Grid environment introduces new challenges that require new approaches.

As part of an investigation of such approaches, we have developed a prototype Grid-enabled framework and runtime. The framework is based on Cactus, a powerful modular toolkit for the construction of parallel solvers for partial differential equations. This framework has been extended with new modules for dynamic data distribution, latency-tolerant communication algorithms, and detection of application slowdown. The runtime exploits services provided by the Globus Toolkit for security, resource discovery, and resource access, and also provides new Resource Locator and Migrator services. We report here on the overall architecture of this Grid-enabled Cactus system and our experiences applying it to a specific challenge problem, namely automated migration of computationally demanding astrophysics computations to "better" resources when currently available resources become overloaded.

The article makes two contributions to the understanding of Grid computing: it describes an architecture for Grid-enabled application frameworks that we believe has broad applicability, and it introduces a novel challenge problem, migration, that we hope will be adopted by others as a test of Grid computing techniques.

The rest of this article is as follows. In Section 2, we provide some background on Cactus and the Globus Toolkit used in our work. In Section 3, we review the nature of the Grid computing problems that we are interested in solving, and in Section 4 we describe our Grid-enabled framework and runtime. Finally, in Sections 5, 6, and 7, we present experimental results, discuss related work, and present our conclusions.

## 2  Background

We provide some background on the Cactus framework and Globus Toolkit that we use in our work.

### *2.1  Cactus*

Originally developed as a framework for the numerical solution of Einstein's Equations [29], Cactus [2, 5] has evolved into a general-purpose, open source, problem solving environment that provides a unified modular and parallel computational framework for scientists and engineers.

The name Cactus comes from its design, which features a central core (or *flesh*) that connects to application modules (or *thorns*) through an extensible interface. Thorns can implement custom-developed scientific or engineering applications, such as computational fluid dynamics, as well as a range of computational capabilities, such as data distribution and checkpointing. An expanding set of Cactus toolkit thorns provides access to many software technologies being developed in the academic research community, such as the Globus Toolkit [17], as described below; HDF5 parallel file I/O; the PETSc scientific computing library [8]; adaptive mesh refinement; web interfaces [28]; and advanced visualization tools.

Cactus runs on many architectures, including uniprocessors, clusters, and supercomputers. Parallelism and portability are achieved by hiding features such as the MPI parallel driver layer, I/O system, and calling interface under a simple abstraction API. These layers are themselves implemented as thorns that can be interchanged and called as desired. The PETSc scientific library has similar concepts of data distribution neutral libraries [7], but Cactus goes further by providing modularity at virtually every level. For example, the abstraction of parallelism allows one to plug in different thorns that implement an MPI-based unigrid domain decomposition, with very general ghost zone capabilities, or an adaptive mesh domain decomposer, or a PVM version of the same kinds of libraries. A properly prepared scientific application thorn will work, without changes, with any of these parallel domain decomposition thorns, or others developed to take advantage of new software or hardware technologies.

The second system that contributes to a Grid-enabled Cactus is MPICH-G [16], an MPI implementation designed to exploit heterogeneous collections of computers. MPICH-G exploits Globus services [17] for resource discovery, authentication, resource allocation, executable staging, startup, management, and control. MPICH-G is distinguished from other efforts concerned with message passing in heterogeneous environments (PACX [19], MetaMPI, STAMPI [22], IMPI [21], MPIconnect [14]) by its tight integration with the popular MPICH implementation of MPI [20] and its use of Globus mechanisms for resource allocation and security.

## 2.2 Globus Toolkit

The Globus Toolkit [17, 18] is a set of services and software libraries that support Grids and Grid applications. The Toolkit includes software for security, information infrastructure, resource management, data management, communication, fault detection, and portability. It is packaged as a set of components that can be used either independently or together to develop useful Grid applications and programming tools. Globus Toolkit components include the Grid Security Infrastructure (GSI), which provides a single-sign-on, run-anywhere authentication service, with support for delegation of credentials to subcomputations, local control over authorization, and mapping from global to local user identities; the Grid Resource Access and Management (GRAM) protocol and service, which provides remote resource allocation and process creation, monitoring, and management services; the Metacomputing Directory Service (MDS) [15], an extensible Grid information service that provides a uniform framework for discovering and accessing system configuration and status information such as compute server configuration, network status, or the locations of replicated datasets; and GridFTP, a high-speed data movement protocol [1]. A variety of higher-level services are implemented in terms of these basic components.

# 3   Cactus and Dynamic Grid Computing

We discuss scenarios that lead Cactus developers and users to require Grid computing technologies, and use these scenarios to arrive at requirements for a Grid-enabled framework and runtime.

Cactus is widely used to perform large-scale simulations in computational astrophysics, for example to study the physics of extreme events such as black hole or neutron star collisions including computing the gravitational wave emissions from such events. The latter calculations are currently of great importance due to the major investments being made in gravitational wave detectors.

The complexity of the Einstein equations that must be solved in numerical relativity means that Cactus simulations can easily consume thousands of floating point operations (flops) per grid point per time step. When applied to the large three-dimensional grids needed for accurate simulations, aggregate computing demands can reach trillions of flops per time step. Thus, Cactus simulations may require weeks of run time on large multiprocessors: this is not a class of simulations that runs effectively on individual workstations.

These tremendous computational requirements have led computational astrophysicists to be aggressive early adopters of parallel computing and major users of supercomputer centers. However, the resources provided by such centers are never sufficient for either the routine computations performed in the course of daily research and development, or the large-scale "reference simulations" performed to probe the limits of resolution, and hence gain new physical knowledge and insights.

In this context, Grid computing becomes attractive as a means both of increasing accessible computing resources and of allowing supercomputer resources to be used in more flexible ways. Grid computing can increase accessible resources by allowing the aggregate workstation and small-cluster computing resources of a collaboration such as the multi-institutional Cactus team to be used for routine calculations. Grid computing can also make supercomputer resources significantly more accessible and useful, for example by hiding the heterogeneity encountered across diverse centers, by allowing migration off overloaded systems, by allowing, in certain circumstances, the coupling of multiple supercomputers to perform extremely large-scale computations.

Table 1 illustrates a scenario that captures the various elements that we believe are required in a complete computing solution for Grid environments. The capabilities required to support this scenario include information service mechanisms able to locate appropriate resources, determine their characteristics, and monitor their state over time; security and resource allocation mechanisms that allow resources to be brought into computations; configurable applications that can be set up to run effectively on various combinations of resources; mechanisms for monitoring application behavior, for example to detect slowdown; and migration mechanisms that allow the set of resources allocated to a computation to be changed over time. Our architecture incorporates all of these capabilities.

**Table 1: Cactus execution scenario in a dynamic Grid environment**

| Time | Activity |
| --- | --- |
| 09:00 | A user poses a computational astrophysics black hole problem that will require around one hundred CPU hours. 50 processors are located at one of the sites participating in the collaboration, and the simulation is started. |
| 09:30 | A 200-processor cluster becomes available. The application is notified and migrates to the new system. |
| 09:50 | The 200-processor cluster becomes unavailable. The system locates 100 processors distributed across three sites connected via high-speed networks, configures Cactus to run on this heterogeneous system, and restarts the simulation. |
| 10:00 | The simulation enters a phase in which it wishes to spawn, every 10 iterations, an analysis task that looks for black hole event horizons in the simulation data. The system locates another 20 processors and brings these into the mix, these processors then working independently of the main simulation |
| 10:15 | Some of the processors allocated at 09:50 become overloaded. The computation is migrated off these processors onto other machines. |
| 10:25 | The computation completes. |

We focus in this article on the particular problem of computation migration, which occurs in several settings in Table 1. In the general case, migration mechanisms must be able to deal with different architecture, operating system, and system resources on the source and target machines, as well as the large-scale transfer of checkpoint data. In addition, when terminating the "old" simulation on the source machine(s) and transferring to a "restarted" simulation on the target machine(s), we may encounter a range of different scenarios:

1. Source and target machines may or may not be available at the same time.

2. The source machines may become unavailable at a known time and allow the "old" simulation to prepare for migration before shutting down gracefully; alternatively, the source machines may shut down without warning, in which case the "old" simulation is terminated with little notice. In a third case, the source machines may be voluntarily abandoned by the "old" simulation due to better resources elsewhere.

3. The new target machines may or may not be known to the old simulation by the time of the migration.

These different cases pose a number of requirements, which need to be addressed by migration framework:

It is necessary to provide external tracking of the status of a simulation to become independent of a sudden simulation shutdown. This requires the simulation to write backup checkpoints at regular intervals. The status tracking includes the location of checkpoints, which are used to recover a simulation and the tracking of the simulation output files.

If a target machine is not immediately available for migration, appropriate storage capacity has to be provided to store the checkpoint files and allow the simulation to "hibernate." Usually compute centers purge large files some time after a simulation finishes. Hence, checkpoint files have to be located and evacuated before they are erased. Due to the size of the transferred checkpoints, the moving of files can take a significant amount of time, and the available network bandwidth must be determined when storing data.

The service that monitors the availability of resources must be implemented externally as well to allow a hibernated simulation to be restarted when a suitable machine configuration becomes online.

The Migrator service is responsible for stopping the application on one set of resources and restarting it at the next sequential point in the calculations on a new set of resources. It is important to make the migration transparent to a scientist who operates a simulation. The migration event appears as an atomic operation to the simulation user. The user can control the migration initiation manually or automatically and define the target machine explicitly or give a preferred machine profile. But he does not have to be concerned about checkpointing the simulation, potentially combining the files if they were written on multiple nodes, transferring the data and restarting the simulation.

The steps that lead to a successful migration can be briefly listed as follows:

1. The migration is initiated. Migration can be initiated in several ways. In Section 4, we describe how migration can be initiated following a contract violation. Alternatively, migration may occur following notification of availability of new compute capacities that provide a better match to a previously defined resource profile. It is important to provide a flexible trigger API to accommodate individual trigger scenarios for the users.

2. The migration framework requests that checkpoint data be written. These checkpoints have to be architecturally independent and will be used to restore the simulation state on a machine whose configuration may differ in architecture, operating system, number of processors, etc. The migration framework informs the external service of the location of these files. If the checkpoint and IO files are in the danger of being purged by the host system, the migration service will copy them to a safe storage location.

3. The checkpoint tracking described above may also be used to recover simulations that have written backup checkpoint files in regular interface and which are terminated or shutdown unexpectedly.

4. The old simulation may send its resource profile to the migration service that characterizes the simulation on the old host. This information can include the memory requirements, FLOP/s, IO patterns etc.

5. The migration framework will make an attempt to restart the simulation on a different machine. The new host may have been obtained from the old simulation or it is looked up by the migration server based on the simulations resource profile. Determining this resource profile and using intelligent logic to rank available machines by efficiency is one of the major goals of this project.

Once a new host is identified, the migration framework will copy all necessary files to the new host and restore the simulation there. Since the new host is not known to the simulation operator, the migration system will inform that the simulation has been relocated to the new host and tell the user how to interact with the new simulation. Alternatively the simulation can be requested to announce itself to a known information service once it has become online again.

# 4   A Grid-Enabled Framework and Runtime

We are developing a software system that addresses the requirements established in the preceding section. As illustrated in Figure 3, this system is intended to support both application-level adaptation and adaptive resource selection; we view these two techniques as equally important means of achieving performance in a dynamic environment.

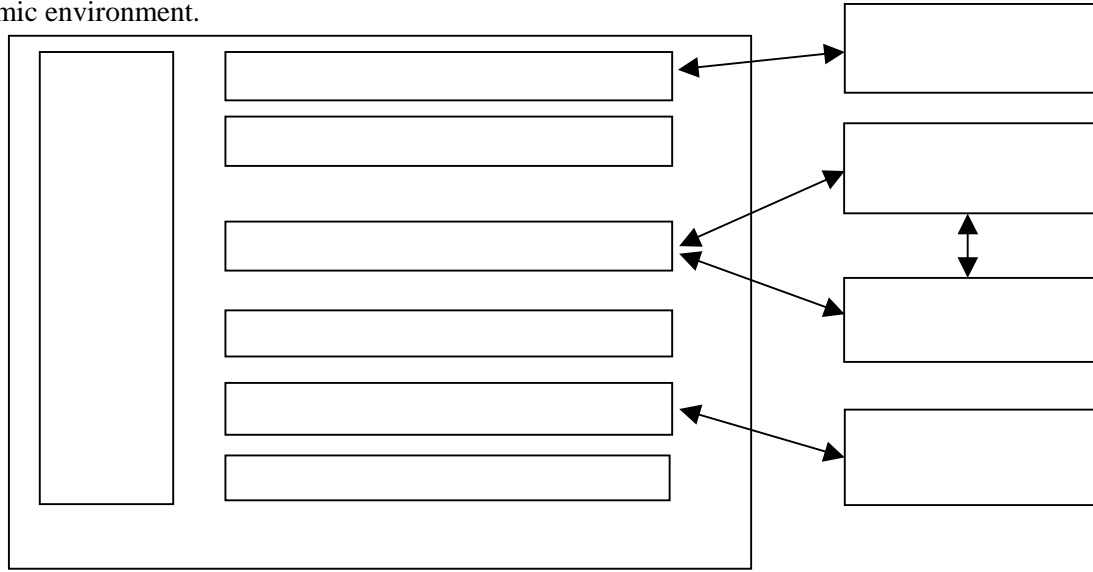

**Figure** 1:  **Cactus Architecture**

The "software system" that we are developing comprises, in fact, two distinct elements: a Grid-enabled framework—the Cactus framework described in Section 2.1, enhanced with a set of thorns (modules) to address adaptation to heterogeneous resource sets—and an adaptive runtime: a set of external services constructed on top of Globus and other services.  The additional Cactus thorns include support for dynamic domain decompositions and adaptive communication schedules [3]; the detection of performance degradation; negotiation with a resource selector; and management of the migration process.  The adaptive runtime includes a resource selector and the migrator that actually performs the transfer of simulation state (accessed as either a checkpoint file or a checkpoint stream) from one resource to the next.  Figure

The "software system" that we are developing comprises, in fact, two distinct elements: a Grid-enabled framework—the Cactus framework described in Section 2.1 and a set of thorns (modules) that extend the Cactus capabilities to provide the migration and resource detection technology.

One set of thorns provides a migration module for any Cactus application, which allows a user to transfer the simulation state by a checkpoint file or checkpoint stream from one resource to the next. The transfer is performed externally from the simulation by the Migrator service, a centralized server process that is contacted by the migrating clients.  Other thorns use adaptive techniques to build on the migration module and add the necessary resource awareness capabilities.  They include a resource selector and mechanisms to measure the simulation code's performance by contract monitoring.  A Migration Logic Manager thorn decides to migrate, depending on unacceptable degradation of performance has occurred on the current resource.

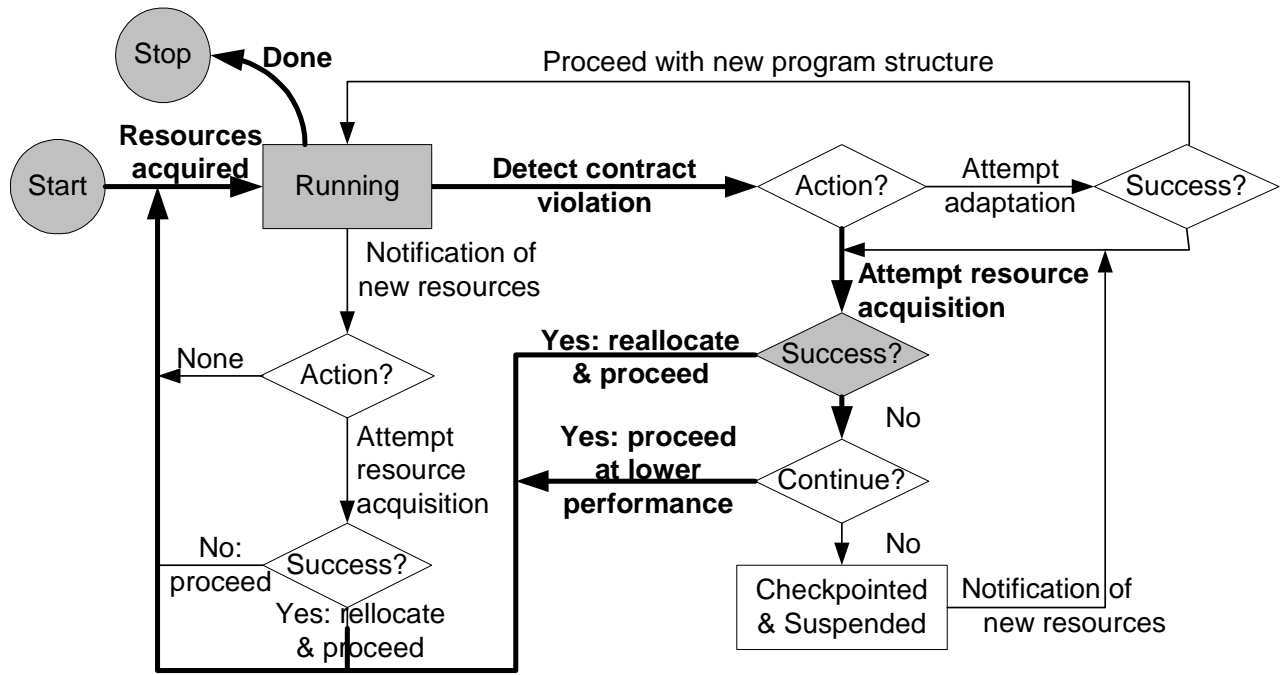

**Figure 2: Flowchart showing the principal decision points that we envision for our Grid-enabled Cactus system. In boldface and shaded are the components discussed in this paper.**

We focus here on the elements in boldface and shaded in the figure, that is, those concerned with resource reallocation as a result of a contract violation detected within the application. These components comprise, specifically, the Worm thorns and the Resource Selector and Migrator services. In brief (see also Figure 3), the Performance Degradation Detection thorn first *determines that a contract violation has occurred* and that resource reallocation may hence be necessary. The Resource Selection Client thorn then *communicates with the Resource Selector*, which *seeks to identify suitable alternative resources* from among those *discovered and characterized by the selector*. If such resources are located, the Migration Logic Manager thorn notifies the Migrator service of the requested migration, which in turn requests a fresh checkpoint and communicate the file's location to the Migrator service to *effect the movement of the computation from one resource to another*. We describe each of these activities in turn.

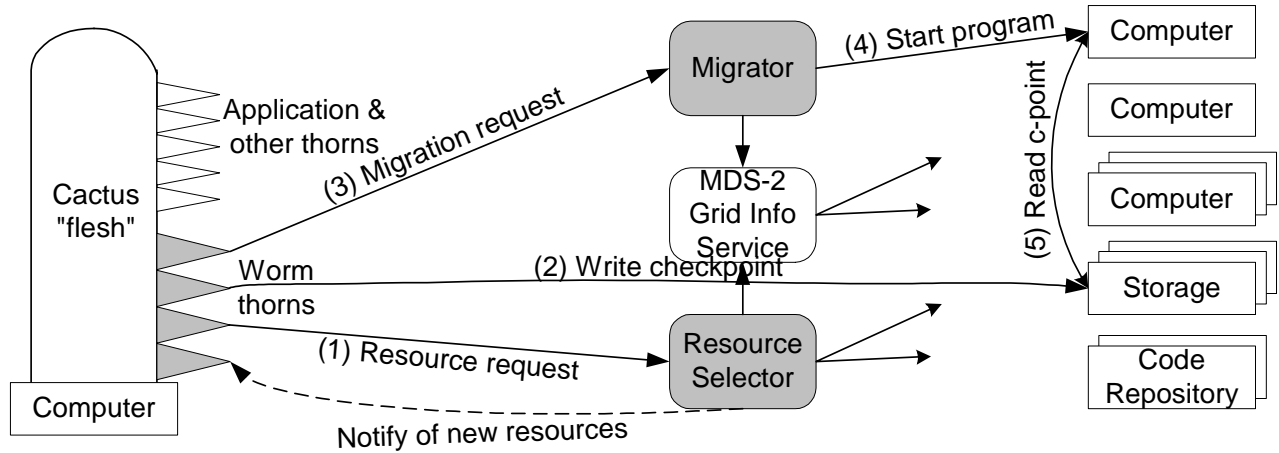

**Figure 3:** Overall architecture of the Grid-enabled Cactus framework, showing in particular the new elements developed in this and related work (shaded), including adaptive thorns, the "Worm" thorns for handling different aspects of migration, and the Resource Selector and Migrator services. Resources are discovered, monitored, and accessed using protocols and libraries provided via the Globus Toolkit. The steps involved in a migration operation are shown, numbered according to the order in which they are performed.

### 4.1 Detecting Contract Violations

The GrADS project has introduced the concept of a performance contract [6]: an agreement between a user and the provider of one or more resources. A contract states that when provided certain *resources* with certain *capabilities* and certain *problem parameters*, the application will achieve a specified sustained level of *measurable performance*. The *resources* can be computational, network, or input/output in nature, and *capabilities* may be expressed in terms of raw computational power, expected load, clock speed, memory, bandwidth, latency, transfer rate, disk capacity, etc. The *problem parameters* could include such things as size of the data, resolution of a simulation, confidence interval, sample size, image resolution, etc. Finally, the *measurable performance* might include things such as flops, transfer rate, number of frames per second rendered, number of iterations per second calculated, etc.

A contract places obligations on both the provider and application. It commits the provider to supply resources with enough capabilities to maintain the promised measurable performance and, at the same time, also commits the application to stay within specified problem parameters.

Providing specific values for the resources, capabilities, problem parameters, and measurable performance parameters instantiates a specific contract. This contract is considered *violated* if, during the execution of the application, the value of any parameter falls outside the specified range. We refer to the online process used to detect such contact violations as contract *monitoring*.

We use a simple contract and contract monitor in our initial implementation. Measurable performance specification is determined dynamically, when the application starts, in terms of iterations calculated per second. The contract is set to specify that the initial performance attained will not degrade more than a certain percentage over the lifetime of the application.

The Performance Degradation Detection thorn monitors the performance of a Cactus application and, if performance degrades below a specified level, informs the Migration Logic Manager which negotiates with the Resource Selector and Migrator services to move the computation to new resources. Detection works as follows. At the end of each time quantum, we compute the execution rate of the application and the performance degradation (if any) relative to the average rate since the computation began. If this degradation is greater than a specified threshold, a violation is registered. If a time quantum was noted as violating the contract, then that time quantum is not used in the calculation of the average rate since the computation began. If more than a specified number of violations occur in a row, then remediation action is attempted. The length of a time quantum, degradation threshold, and number of consecutive violations required before action constitute parameters, that in our implementation can be changed at runtime by contacting the application through an HTTP interface (provided by a separate Cactus thorn).

Remediation involves first interacting with the Resource Selector service to determine if any resources are available that can offer better performance than those currently in use. Assuming that a suitable resource is located, the Migration Logic Manager instructs the Migration Module to migrate. The simulation state is written to a checkpoint, which is stored to stable storage. The Migration Module then contacts the external Migrator service to request that the computation be restarted on the new resource. We describe the workings of the Migrator in more detail below.

## 4.2 Resource Selection

The Resource Selector is an independent service responsible for resource discovery and selection based on application-supplied criteria. This service discovers and keeps track of available resources, using techniques that we describe here, and responses to requests from applications (or other Grid services). These requests are communicated via a synchronous, query-response protocol based on the ClassAds syntax [24] defined within the Condor project [23]. A request is a ClassAd representing the criteria that should be used to select resources; the response indicates either success or failure, with a specification for the matching resources being supplied in the event of success. We plan in the future to extend the protocol to support asynchronous notification upon detection of new resources.

We first describe the techniques used to discover available resources, and then describe the selection process. Resource discovery exploits services provided by the Globus Toolkit's MDS [12], which defines two key protocols: the GRid Registration Protocol (GRRP), which a resource uses to register with an aggregate directory, and the GRid Information Protocol (GRIP), which an aggregate directory or user uses to enquire about the status of a resource. The Globus Toolkit also provides a simple aggregate directory, which caches information about registered resources.

Our Resource Selector uses MDS as follows. When it starts up, and periodically thereafter, it queries appropriate aggregate directory(s) to locate any potentially interesting resources. Having obtained the names and gross characteristics of these resources, the Resource Selector then uses GRIP to contact them directly and obtain detailed information. The Resource Selector then continues to monitor those resources over time, via periodic queries with frequencies based on time-to-live information. In the future, we plan also to explore the use of asynchronous notifications and adaptive query strategies based on expected application queries.

Our selection process is based on the Condor "matchmaking" algorithm, which in its basic form takes two ClassAds and evaluates one with respect to the other. A ClassAd is a set of expressions that must all evaluate to true in order for a match to succeed. Expressions in one can be evaluated using property values from the other. For example, the expression "`other.size > 3`" in one ClassAd evaluates to **true** if the other ClassAd has an attribute named `size` and the value of that attribute is (or evaluates to) an integer greater than three. A ClassAd can also include a rank expression that evaluates to a numeric value representing the quality of the match.

Our selection process operates as follows. When a request is received (in the form of a ClassAd, as noted earlier), the Resource Selector invokes the matchmaking algorithm against ClassAds representing available resources, and returns the "best" match (if any), as determined by the computed ranks.

This leaves open the question of how exactly the matching of a request with available resources is performed. There are two issues of concern here:

1. A request will ask for multiple computers, and in principle any combination of available computers could match the request. Yet it is clearly impractical to try all possible combinations.
2. The ClassAd language has limited expressivity: for example, it does not contain iterators, so we cannot express constraints such as "the total memory of all processors should be at least 10 GB."

Our current approach is to view available resources as being grouped into a number of "cliques." (For simplicity, we currently define these cliques manually, but they can also be constructed automatically, e.g., on the basis of domain name or network connectivity as measured by network performance experiments.) The ResourceSelector constructs, and maintains current, one ClassAd for each clique, with this ClassAd containing expressions specifying various derived values representing attributes of the computers in the clique, such as number of processors, minimum link bandwidth, bisection bandwidth, and so forth. This use of derived values allows us to circumvent the limitations of the ClassAd language, but does constrain the types of matching that can be performed. We then match each request against all available cliques, and select the best on the basis of rank.

We can imagine a variety of alternative approaches. For example, we could substitute an alternative matchmaking algorithm that uses a greedy strategy to build a set of candidate resources. Or, we could use some other heuristic (e.g., genetic algorithms) to identify good candidates. We may investigate these in the future.

In our current work, client requests specify relatively simple criteria for resource selection, such as operating system version, minimum memory, and minimum bandwidth. In the future, we plan to explore the representation of more complex performance models [27]. Ideally, we would like to be able to encode accurate performance models in the request ClassAd, thus achieving a declarative specification of resource requirements that can be used in the Resource Selector. Figure 4 shows the ClassAds format. Note that "Type," "Owner," "requirements," "rank," and "other" are predefined keywords in the ClassAds parser; however, the "RequiredDomains" variable is user defined. The "other" keyword refers to the advertisement ClassAd against which this request ClassAd is compared for a potential match.

```
[
  Type="request";
  Owner="dangulo";
  RequiredDomains={"cs.uiuc.edu", "ucsd.edu"};
  requirements= "other.opSys=='LINUX' &
          other.minMemSize> (100G/other.CPUCount) &&
          Include(other.domains, RequiredDomains)
         ";
  Rank= other.minCPUSpeed * other.CPUCount / (other.maxCPULoad+1);
]
```

**Figure** 4: **ClassAds Format.** Type, Owner, requirements, rank, **and** other **are predefined keywords in the ClassAds parser; however, the** RequiredDomains **variable is user defined. The** other **keyword refers to the advertisement ClassAd against which this request ClassAd is compared for a potential match.**

### 4.3 Migration

Let us assume that a contract violation has been detected and the resource selector has identified appropriate alternative resources. We can therefore proceed to migrate the application, as follows:

- The Migration Logic Manager thorn instructs Cactus to checkpoint the application state to stable storage, either locally or (perhaps more efficiently) to a remote location. The Migrator is informed of the current location of the checkpoint files.

- The Migration Module thorn communicates with the external Migrator service, informing it of the current simulation location and the new target host and tells it to restart the application on the new resources. The Migrator Service stages all required files and executables to the new machine.

- The Migrator Service uses the GRAM protocol to request that resources are allocated and the Cactus application is started on the remote resource.

- The Cactus application is restarted, reads the previously generated architecturally independent checkpoint file and continues execution.

We note that as Globus services are used for these activities, all are authenticated via GSI mechanisms and hence we are protected against certain attacks, such as someone submitting a bogus migration request to the Migrator Service.

## 5  Experiences

We have experimented with various versions of the system described above on Grid testbeds operated by the Egrid consortium [4] and the GrADS project. We describe here our experiences on the GrADS testbed, which comprises workstation clusters at universities across the U.S., including U.Chicago, UIUC, UCSD, Rice University, and USC/ISI.

In each of these experiments, we ran a simulation on a SUSE Linux 7.0 machine at the University of Chicago. At a certain time, we introduced an artificial load by starting another computational task on the same machine. The Degradation Detection parameters were set up to record a Contract Violation when the degradation exceeded 10 percent of computation time. Other parameters were set to initiate migration when three successive Contract Violations were encountered. The Resource Selector was configured to select from among the various resources in the GrADS testbed.

The results shown in Figure 5 are typical. Initially, the simulation runs normally. At time step 7, an artificial load is introduced onto the UC machine, and computational degradation was detected at time steps 8, 9, and 10. Each of these degradation values was sufficiently high to classify them as Contract Violations. These three successive contract violations motivate migration of the calculation process to a set of resources where computational throughput exceeds the degraded throughput. As seen from figure 7, the throughput on the new set of resources (at UIUC) is much improved over the degraded throughput obtained on the initial resources (at UC), although the throughput is not as high as that obtained on the initial unencumbered resources. This is what was expected, as otherwise the resources at UIUC would have been selected to run on initially.

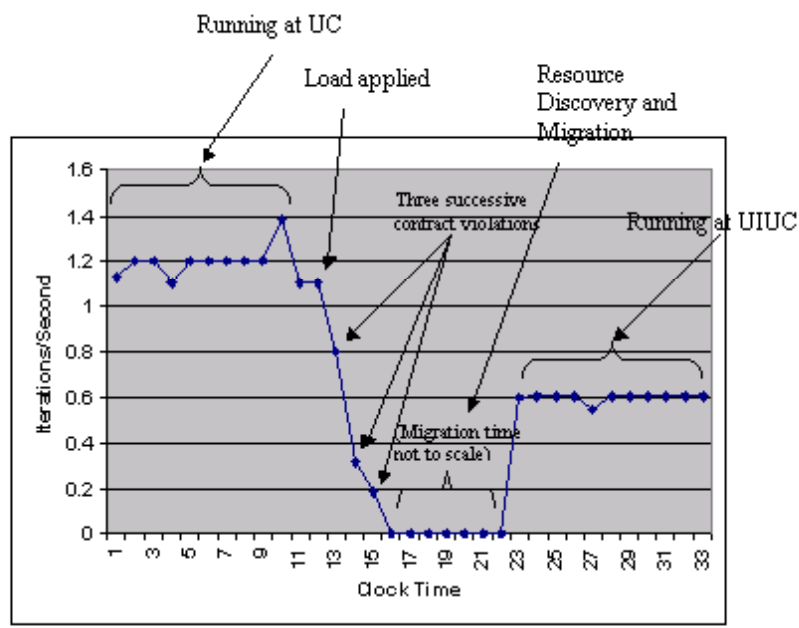

**Figure 5: Measured Throughput During Migration**

Migration of the calculation process is an expensive operation. Currently, to migrate checkpoint data whose size is $O$(96) MB takes $O$(200) seconds. The time is excessive because of the number and types of operations performed in our prototype: each migration operation involves not only communication with the Resource Selector but also the checkpoint, migration of the checkpoint files to the Migrator Service, and subsequent movement of the checkpoint files from the Migrator Service to the eventual destination. Hence, the computation state is moved to or from disk *six* times, when in principle a single memory-to-memory transfer over the network would suffice.

# 6 Related Work

This work has been performed in the context of the Grid Application Development Software (GrADS) project, a multi-institutional NSF-funded research effort that is investigating new approaches to program development process appropriate for Grid environments. Many of the concepts described here have been developed in the course of discussions with GrADS colleagues. Other papers describe the overall GrADS project approach [10], and results obtained in a related project focused on Grid execution of ScaLAPACK [26].

Many other projects have addressed specific issues discussed in this project, including resource selection (e.g., [9, 11, 30]), mapping (e.g., [31]), and process migration (e.g., [13, 25, 32]). We do not claim innovation in these areas, but instead emphasize the merits of our overall architecture and, in particular, the techniques used to integrate adaptive mechanisms into the Cactus architecture.

# 7 Summary and Further Work

We have described a Grid-enabled computational framework (Cactus) and associated Grid services (Resource Selector, Migrator) that together are capable of automated migration to "better" resources following contract violation in a wide area Grid environment. As we have emphasized throughout the paper, the framework and services are relatively simple proof-of-principle prototypes that rely on numerous simplifying assumptions. However, they do provide a framework within which much useful further work can be performed. In addition, we believe that this "Cactus Worm" represents a wonderful challenge problem for Grid computing.

We have referred to several topics for future work in the text; we summarize these and others here. We wish to investigate how to automate two of the more challenging tasks facing the Cactus programmer, namely the insertion of contract violation code and the generation of performance models. Fortunately, several of our colleagues in the GrADS project are investigating exactly those issues. We plan also to pursue further the nature of the resource selection process, with the goal of understanding tradeoffs between complexity of selection criteria, resource discovery techniques, and effectiveness of selection. We also wish to develop more accurate performance models that take into account, for example, the cost of migration. Finally, we will address the performance of the migration process itself.

## Acknowledgements

This work was supported in part by the NSF-funded Grid Application Development Software project under Grant No. 9975020. We are grateful to our GrADS project colleagues for discussions on the topics discussed here.

## References

1. Allcock, B., Bester, J., Bresnahan, J., Chervenak, A.L., Foster, I., Kesselman, C., Meder, S., Nefedova, V., Quesnel, D. and Tuecke, S., Secure, Efficient Data Transport and Replica Management for High-Performance Data-Intensive Computing. in *Mass Storage Conference*, (2001).
2. Allen, G., Benger, W., Goodale, T., Hege, H., Lanfermann, G., Merzky, A., Radke, T., Seidel, E. and Shalf, J. Cactus Tools for Grid Applications". *Cluster Computing*.
3. Allen, G., Dramlitsch, T., Foster, I., Goodale, T., Karonis, N., Ripeanu, M., Seidel, E. and Toonen, B., Supporting Efficient Execution in Heterogeneous Distributed Computing Environments with Cactus and Globus. in *SC'2001*, (2001), ACM Press.
4. Allen, G., Dramlitsch, T., Goodale, T., Lanfermann, G., Radke, T., Seidel, E., Kielmann, T., Verstoep, K., Balaton, Z., Kacsuk, P., Szalai, F., Gehring, J., Keller, A., Streit, A., Matyska, L., Ruda, M., Krenek, A., Frese, H., Knipp, H., Merzky, A., Reinefeld, A., Schintke, F., Ludwiczak, B., Nabrzyski, J., Pukacki, J., Kersken, H.-P. and Russell, M., Early experiences with the Egrid testbed. in *IEEE International Symposium on Cluster Computing and the Grid*, (2001).
5. Allen, G., Goodale, T., Lanfermann, G., Seidel, E., Benger, W., Hege, H.-C., Merzky, A., Mass\'o, J., Radke, T. and Shalf, J. Solving Einstein's Equation on Supercomputers. *IEEE Computer* (December). 52-59.
6. Aydt, R., Mendes, C., Reed, D. and Vraalsen, F. Specifying and Monitoring GrADS Contracts, UIUC, 2001.
7. Balay, S., Gropp, W.D., McInnes, L.C. and Smith, B.F. Efficient Management of Parallelism in Object Oriented Numerical Software Libraries. in Langtangen, E.A.a.A.M.B.a.H.P. ed. *Modern Software Tools in Scientific Computing*, 1997, 163--202.
8. Balay, S., Gropp, W.D., McInnes, L.C. and Smith, B.F. PETSc 2.0 Users Manual. (ANL-95/11 - Revision 2.0.22).
9. Berman, F. High-Performance Schedulers. in Kesselman, C. ed. *The Grid: Blueprint for a New Computing Infrastructure*, Morgan Kaufmann, 1999, 279-309.
10. Berman, F., Chien, A., Cooper, K., Dongarra, J., Foster, I., Gannon, D., Johnsson, L., Kennedy, K., Kesselman, C., Mellor-Crummey, J., Reed, D., Torczon, L. and Wolski, R. The GrADS Project: Software Support for High-Level Grid Application Development. *International Journal of Supercomputer Applications*.

11. Berman, F., Wolski, R., Figueira, S., Schopf, J. and Shao, G. Application-Level Scheduling on Distributed Heterogeneous Networks. in *Proc. Supercomputing '96*, 1996.
12. Czajkowski, K., Fitzgerald, S., Foster, I. and Kesselman, C., Grid Information Services for Distributed Resource Sharing. in *IEEE International Symposium on High Performance Distributed Computing*, (2001), IEEE Press.
13. Douglis, F. and Ousterhout, J. Transparent Process Migration: Design Alternatives and the Sprite Implementation. *Software Practice and Experience*, *21* (8). 757--785.
14. Fagg, G., Dongarra, J. and Geist, A. PVMPI Provides Interoperability Between MPI Implementations. in *Proc. 8th SIAM Conf. on Parallel Processing*, 1997.
15. Fitzgerald, S., Foster, I., Kesselman, C., Laszewski, G.v., Smith, W. and Tuecke, S. A Directory Service for Configuring High-performance Distributed Computations. in *Proc. 6th IEEE Symp. on High Performance Distributed Computing*, 1997, 365--375.
16. Foster, I. and Karonis, N. A Grid-Enabled MPI: Message Passing in Heterogeneous Distributed Computing Systems. in *Proc. SC'98*, 1998.
17. Foster, I. and Kesselman, C. Globus: A Toolkit-Based Grid Architecture. in Kesselman, C. ed. *The Grid: Blueprint for a New Computing Infrastructure*, Morgan Kaufmann, 1999, 259-278.
18. Foster, I., Kesselman, C. and Tuecke, S. The Anatomy of the Grid: Enabling Scalable Virtual Organizations. *Intl. J. Supercomputer Applications*, *(to appear)*.
19. Gabriel, E., Resch, M., Beisel, T. and Keller, R. Distributed Computing in a Heterogenous Computing Environment. in *Proc. EuroPVM/MPI'98*, 1998.
20. Gropp, W., Lusk, E., Doss, N. and Skjellum, A. A High-Performance, Portable Implementation of the MPI Message Passing Interface Standard. *Parallel Computing*, *22*. 789--828.
21. IMPI Steering Committee IMPI - Interoperable Message-Passing Interface.
22. Kimura, T. and Takemiya, H. Local Area Metacomputing for Multidisciplinary Problems: A Case Study for Fluid/Structure Coupled Simulation. in *Proc. Intl. Conf. on Supercomputing*, 1998, 145--156.
23. Livny, M. High-Throughput Resource Management. in Kesselman, C. ed. *The Grid: Blueprint for a New Computing Infrastructure*, Morgan Kaufmann, 1999, 311-337.
24. Livny, M., Matchmaking: Distributed Resource Management for High Throughput Computing. in *IEEE International Symposium on High Performance Distributed Computing*, (1998), IEEE Press.
25. Miller, M.L.P.a.B.P., Process migration in DEMOS/MP. in *Ninth ACM Symposium on Operating System Principles*, (1983), 110-119.
26. Petitet, A., Blackford, S., Dongarra, J., Ellis, B., Graham Fagg, Roche, K. and Vadhiyar, S. Numerical Libraries And The Grid: The GrADS Experiments With ScaLAPACK. *International Journal of Supercomputer Applications*.
27. Ripeanu, M., Iamnitchi, A. and Foster, I. Performance Predictions for a Numerical Relativity Package in Grid Environments. *International Journal of Scientific Applications*, *14* (4).
28. Russell, M., Allen, G., Goodale, T., Foster, I., .Suen, W., Seidel, E., Novotny, J. and Daues, G., The Astrophysics Simulation Collaboratory: A Science Portal Enabling Community Software Development. in *IEEE International Symposium on High Performance Distributed Computing*, (2001), IEEE Press.
29. Seidel, E. and Suen, W.M. J. Comp. Appl. Math., *109*. 493-525.
30. Subhlok, J., Lieu, P. and Lowekamp, B. Automatic Node Selection for High Performance Applications on Networks. in *Proceedings of the Seventh ACM SIGPLAN Symposium on the Principles and Practice of Parallel Programming (PPoPP'99)*, 1999, 163--172.
31. Subhlok, J., O'Hallaron, D., Gross, T., Dinda, P. and Webb, J. Communication and Memory Requirements as the Basis for Mapping Task and Data Parallel Programs. in *Proceedings of Supercomputing '94*, Washington, DC, 1994, 330-339.
32. Theimer, M.M. and Hayes, B., Heterogeneous Process Migration by Recompilation. in *11th International Conference on Distributed Computing Systems*, (1991), 18-25.

Page: 12